# Multidimensional Social Network in the Social Recommender System[1]

Przemysław Kazienko, Katarzyna Musiał and Tomasz Kajdanowicz

*Abstract* — All online sharing systems gather data that reflects users' collective behaviour and their shared activities. This data can be used to extract different kinds of relationships, which can be grouped into layers, and which are basic components of the multidimensional social network proposed in the paper. The layers are created on the basis of two types of relations between humans, i.e. direct and object-based ones which respectively correspond to either social or semantic links between individuals. For better understanding of the complexity of the social network structure, layers and their profiles were identified and studied on two, spanned in time, snapshots of the *Flickr* population. Additionally, for each layer, a separate strength measure was proposed. The experiments on the *Flickr* photo sharing system revealed that the relationships between users result either from semantic links between objects they operate on or from social connections of these users. Moreover, the density of the social network increases in time.

The second part of the study is devoted to building a social recommender system that supports the creation of new relations between users in a multimedia sharing system. Its main goal is to generate personalized suggestions that are continuously adapted to users' needs depending on the personal weights assigned to each layer in the multidimensional social network. The conducted experiments confirmed the usefulness of the proposed model.

## I. INTRODUCTION

WEB 1.0 and some internet services like email systems enable to extract and analyze social networks based on data about activities of single user [9]. In turn, Web 2.0 applications facilitate collaborative actions of users in which informal, dynamic groups of people cooperate or share common interests with one another. Recently, the multimedia sharing systems (MSS) like *Flickr* or *YouTube*, which are typical examples of Web 2.0 systems, successfully attract more and more users who can share their multimedia content such as photos, videos, animations, etc. The MSS users have also the opportunity to make public and share content they provide as well as express their opinions about multimedia objects (MOs) authored and maintained by other users. Each multimedia object published in MSS can be tagged by its author. In other words, users can describe their MOs with one or more short phrases, which are most meaningful for the authors and which usually express the MO content in the textual form. Users can also comment the items added by others, include them to their favourites, etc. A comment on a photo, which is made public in MSS, is a sign of similar interests between the author and the commentator. These similar or shared user activities related to MOs reflect indirect, object-based contacts between users. Furthermore, users may set up new, direct relationships with other system users by direct enumeration of their friends or acquaintances.. The users can also establish groups of collective interests. Thus, MSS users interact, collaborate and influence one another and in this way get into conscious or unconscious relationships and form a kind of self-organising social community [13]. Additionally, most of these relations are visible for all system users, which increases their sense of community.

Users act using semantic premises and relationships between multimedia objects they are interested in. Nevertheless, they simultaneously exploit social links to people they know or like.

Overall, these relationships can be relatively easily extracted from the data about user activities. The MSS users together with their interpersonal direct and indirect relationships can be treated as one heterogeneous, multidimensional social network called also multi-layered social network [24] or multi-relational social network [37].

The main goal of the paper is to analyze profiles of different layers within the multidimensional social network extracted from the data available in the photo sharing system. These layers can reflect both semantic and social inspirations of user activities. The former result from recent user needs and interests whereas the latter correspond to users' acquaintances and social preferences. Finally, this multidimensional social network is used in recommender system to suggest one human being to another and in consequence to expand the human community. It mainly makes use of relationships that are not explicitly visible for users, because most relationships come

Manuscript received on        .
The work was supported by The Polish Ministry of Science and Higher Education, the development project, 2009-11 and the research project, 2010-13.
P. Kazienko is with the Wroclaw University of Technology, Wroclaw, 50-370, POLAND (e-mail: kazienko@pwr.wroc.pl).
K. Musiał is with the Wroclaw University of Technology, Wroclaw, 50-370, POLAND and the Bournemouth University, Poole, Dorset, BH12 5BB, UNITED KINGDOM (e-mail: katarzyna.musial@pwr.wroc.pl).
T. Kajdanowicz is with the Wroclaw University of Technology, Wroclaw, 50-370, POLAND (e-mail: tomasz.kajdanowicz @pwr.wroc.pl).
[1] This is not the final version of this paper. You can find the final version on the publisher web page.

from indirect connections via MOs rather than from direct links.

## II. RELATED WORK

Recommender systems have become an important part of the web sites; the vast number of them is applied to e-commerce. They help people to make decision, what items to buy [22], [28], which news to read [47] or which movie to watch [10]. Recommender systems are especially useful in environments with information overload since they cope with selection of a small subset of items that appear to fit the user's preferences [2], [35], [45], [50]. Furthermore, these systems enable to maintain the loyalty of the customers and increase the sales [22].

In general, four categories of recommender systems can be enumerated: demographic filtering, collaborative filtering [44], content-based filtering, as well as their hybrid fusions [2], [35]. Demographic filtering approaches use descriptions of users to extract the relationship between an item and groups of persons that find it interesting [29]. Users are classified based on personal (demographic) data provided by them during the registration process. Alternatively, the same data can be extracted from purchasing history, survey responses, etc. Each item (a web page or a product) is assigned to one or more classes with a certain weight. Similarly, the user profile is matched against the classes and the items related to the closest one are recommended.

The collaborative filtering technique relies on opinions about items delivered by other users. The system recommends products or other items that have been highly evaluated by other people, whose ratings and tastes are similar to the preferences of the user who will receive recommendation [2], [16], [47]. There are two main variants of collaborative filtering in which either *k*-nearest neighbours or a whole, previously extracted nearest neighbourhood are used.

In the content-based filtering the items recommended to the user are similar to the items that user had liked previously [36].

As regards the hybrid methods, some of the three abovementioned basic approaches are combined [19] [22].

Usually, recommender systems are used to suggest different products or services [8] [20]. However, the new application domain for recommender systems are multimedia sharing systems like *Flickr* or *YouTube* that have rapidly developed in the web and usually support thousands or even millions of their users. The main goal of a recommender system in this case is either to suggest new multimedia content [36] or to find some other users who could be interesting for a given one and in consequence to help a user to establish new interpersonal relationships [38]. A recommender system that suggests multimedia objects (MOs) based on social similarity of ontologies maintained separately for each user and each multimedia object was presented in [33] and [36].

People who interact with one another, share common activities or even possess similar demographic profiles can form a social network. Overall, the concept of a social network is quite simple and can be described as a finite set of individuals, by sociologists called actors, who are the nodes of the network, and ties that are the linkages between them [1], [11], [14], [17], [48]. In other words, social network indicates the ways in which actors are related. Tie between actors can be maintained according to either one or several relations [14] and these relations may be either weighted or unweighted; the former can be treated as fuzzy [49]. Moreover, the network gives egos (focal actors) access not only to their alters (people who are directly connected with ego), but also to alters of their alters [14], also called "friends of friends". Nodes of a social network are not independent beings. Some characteristics that describe members of a network can be defined, e.g. demographic and interest data about people. However, none of social network analysis (SNA) methods samples the individuals independently. Actors are connected via relationships and such structure is studied. Several measurements can be applied to investigate the number and the quality of the relationships within the network. The crucial techniques currently used to identify the structure of a social network are: full network method, snowball method, and ego centric methods [17]. The analysis of various structural metrics for social networks together with respect to their application in recommender systems was presented in [40]. On the other hand, Perugini et al. utilized recommendations as connections between users and studied these relationships [41].

The continuously increasing popularity of the World Wide Web and the Internet caused that more and more various types of services, where people can exchange information, are available. People who use these services have created a new kind of virtual societies called online social networks [18] a.k.a. web-based social networks [15], computer-supported social networks, virtual communities or social network sites [5]. Although the basic concept of online social networks is similar to the regular one, their characteristics differ. One of the features that distinguishes regular social network from the existing in the Internet is the relative high easiness of gathering data about communication or common activities and its further processing. The global network provides a vast amount of diverse data useful for social network analysis, e.g. for the estimation of the user social position [23] or finding significant individuals or objects [6]. Internet-based social networks can be either directly maintained by dedicated web systems like Facebook [31], Friendster [4], MySpace [3], and LinkedIn [7] or extracted from data about user activities in the communication networks like e-mails, chats, blogs, homepages connected by hyperlinks [1], etc. Some researchers identify the communities within the Web using link topology [12], while others analyze the emails to discover the social network [9].

The multimedia sharing systems like *Flickr*, *YouTube* or *Broadcaster.com* can be also seen as social networks, where relations among users are extracted from common communication or activities. Such systems enable user to upload and manage multimedia content such as photos, videos, animations, commonly called multimedia objects (MOs). Each of the multimedia objects can be tagged by the author. It means that user can describe MO with one or more short phrases that usually denote the content of this element. Tags used by members can be the basis for creating the social network based on tagging, in which the members are the nodes of the network and the relationship between two members exist if both of them have used at least one common tag to describe their multimedia objects. Simultaneously, they interact, collaborate and influence one another. Users can not only tag the items they have published but also comment the items added by others, include them to their favourites, etc. Additionally, users have the opportunity to set up new, direct relationships with other system users.

*Flickr*, which is analyzed in this paper, has already been the subject of some studies but to date it has rather been treated as a social tagging system that enables users to mark their pictures with tags and then share these tags with other users [27], [30], [39], [42]. The new human relations emerge from users' common tags [32]. Separately, direct links between users (contacts) were studied in [34] with respect to the growth of the communities. However, the relations can be extracted also from other data available in the system and this is the basis to treat online publishing systems as multidimensional social networks, in which there may be more than one kind of relation between two users [47]. The recommender system for collective tag suggestions for *Flickr* was proposed in [43].

Some preliminary research on multidimensional social networks were presented by authors in [37], where nine different types of relations between users were extracted from *Flickr*, i.e. contact lists, tags, groups of items and their authors, favourite pictures and comments to pictures. Some of them like favourites and opinions were split into three separate layers, e.g. author – commentator, commentator – author, commentator – commentator. Tojo et al. analyzed multidimensionality of the *Flickr* social network from another perspective – as the homogeneous coherent ontology [46].

The concept of human recommendation in *Flickr* was first proposed by authors in [38]. In this paper, firstly social layers in *Flickr* were defined and analyzed (Sec. III) as well as experimentally compared with each other using two data sets, for two separate years: 2007 and 2008 (Sec.IV). Two additional layers were identified and investigated compared to those from [25] and [38]. Next, the detailed and deeper insight to both the recommendation concept (Sec. V) and the experiments on the real system (Sec. VI) were presented.

### III. MULTIDIMENSIONAL SOCIAL NETWORK

All direct links or cooperation-based connections are based on individual features of MSS available for users. The set of linkages $L$ is derived directly from data about user activities such as tagging, user groups, comments to multimedia objects (MOs), favourite MOs or contact lists. Each of the activity may bind users in a different way so it forms a relation of different kind. The tie $l_{ij}=(u_i,u_j)\in L$ exists if and only if there exists at least one relation of any type. Thus, every tie $l_{ij}$ can consist of one or more relations $r_{ij}$ which are connections of the specific type from $u_i$ to $u_j$. Three kinds of relations can be distinguished:

1. *Direct intentional relation* $r_{ij}$ from user $u_i$ to $u_j$ exists if user $u_i$ directly points to $u_j$, e.g. by adding $u_j$ to the $u_i$'s contact list.
2. *Object-based relation with equal roles* $r_{ij}$ means that users $u_i$ and $u_j$ meet each other through the meeting object and their role $a$ towards this object remains the same. Usually, they share the same activity in the system, e.g. two users comment the same picture, both add the same object to their favourites or both assign the same tags to describe their photos (Figure 1a).
3. *Object-based relation with different roles* $r_{ij}^{ab}$, $r_{ji}^{ba}$ – is the relation between two users $u_i$ and $u_j$ that are connected through the meeting object but their roles $a$ and $b$ are different, e.g. user $u_i$ comments a photo (role $a$ – commentator) that was published by user $u_j$ (role $b$ – author) (Figure 1b). A non zero relation $r_{ij}^{ab}$ entails the non zero relation $r_{ji}^{ba}$.

Note that object-based, indirect relations are usually relevant to interests common for two or more users whereas direct intentional relations reflect mutual acquaintance. It means that object-based relations are more thematic while direct intentional are more social. All of them enable to create the strongly related semantic group of users.

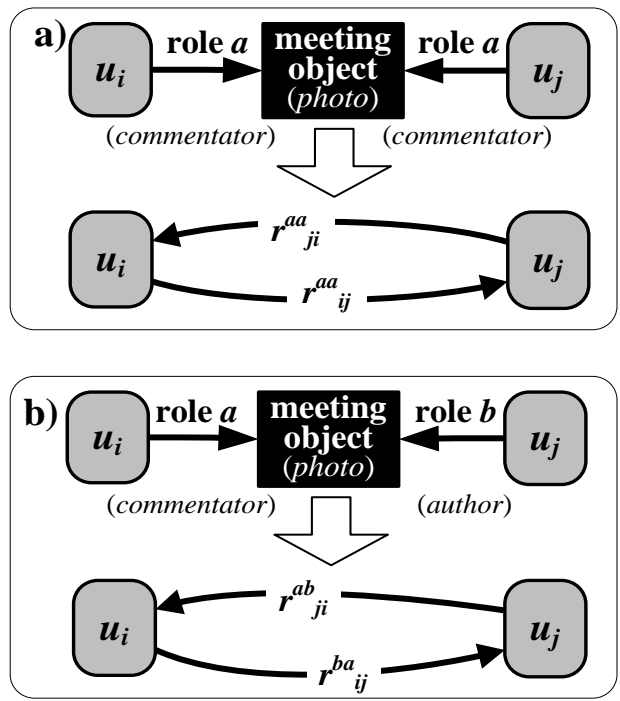

Fig. 1. The object-based relation with equal role – commentator (a), and with two different roles: commentator and author (b)

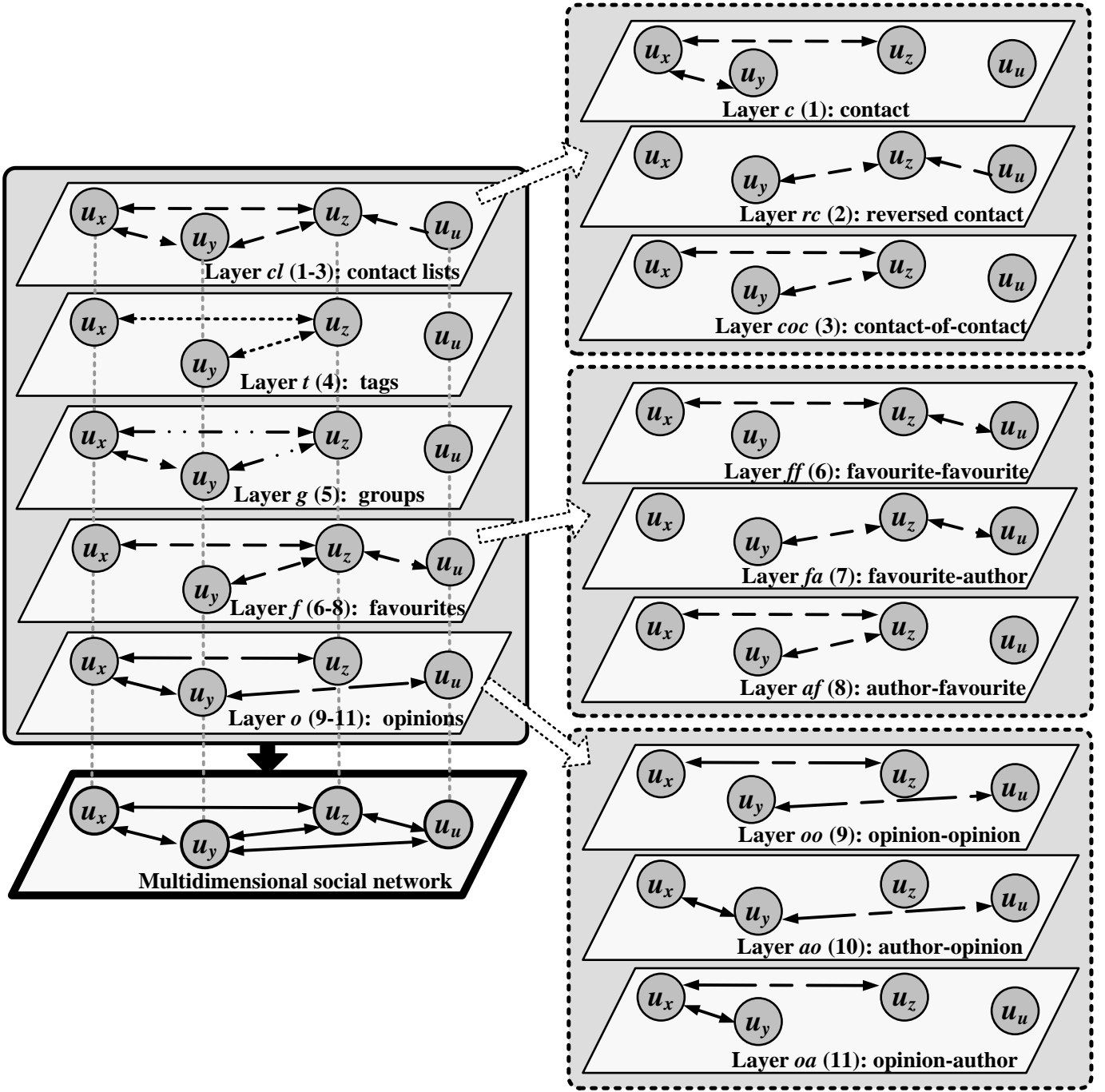

Fig. 2. The relation layers in Flickr

**Definition 1**. *A multidimensional social network MSN is a tuple (U,L), where U is the finite set of non-anonymous user accounts registered in the multimedia sharing system MSS. A single tie i.e. linkage $l_{ij}=(u_i,u_j)\in L$, which denotes the connection from user $u_i\in U$ to user $u_j\in U$, exists if and only if there exists direct intentional link from user $u_i$ to $u_j$, or if there is any common activity of both $u_i$ and $u_j$. The last case results in existence of two ties $(u_i,u_j)\in L$ and $(u_j,u_i)\in L$. The set U must not contain isolated users, i.e. $\forall u_i\in U\ \exists u_j\in U, i\neq j\ ((u_i,u_j)\in L \lor (u_j,u_i)\in L)$.*

Multidimensional social networks may also be called multi-layered social networks [24] or multi-relational social network [37]. A typical representation of multidimensional social network is multigraph.

### A. Relation Layers in the Photo Sharing System

The concepts of social network and ties that aggregate different types of relations were applied to the *Flickr* photo sharing system, in which MOs are photos. Users can publish their pictures in *Flickr*, mark them with tags, create groups and attach their photos to them, build their own lists of favourite photos published by others, maintain contact lists linking to their acquaintances as well as comment photos authored by others. All these activities reflect common interests or acquaintances between users and enable to create the multidimensional social network.

During the research eleven types of relations were identified in *Flickr*: relations based on contact lists – $R^c$, $R^{rc}$, $R^{coc}$, shared tags used by more than one user – $R^t$, user groups – $R^g$, photos added by users to their favourites – $R^{ff}$, $R^{fa}$, $R^{af}$, and opinions about pictures created by users – $R^{oo}$, $R^{oa}$, $R^{ao}$. Relations based on contact lists ($R^c, R^{rc}, R^{coc}$) represent direct intentional relations. Tag-based ($R^t$), group-based ($R^g$), favourite-favourite ($R^{ff}$), and opinion-opinion ($R^{oo}$) relations are typical object-based relations with equal roles, whereas favourite-author ($R^{fa}$), author-favourite ($R^{af}$), opinion-author ($R^{oa}$), and author-opinion ($R^{ao}$) are object-based relations with different roles. All these relations correspond to eleven separate layers in one multidimensional social network, Figure 2.

Each relation is extracted from the data about user behavior and can have assigned either unary (1) or real values. These values express the strength of the relation and are specific for each layer. Overall, the greater user $u_i$'s activity towards user $u_j$ among all activities of $u_i$, the stronger the relationship from $u_i$ to $u_j$.

### B. Direct and Indirect User-Based Relations Derived from Contact Lists

The information about user $u_i$'s relations based on contacts is derived directly from $u_i$'s contact list ($CL_i$), Figure 3. The relation $r_{ij}^c$ from user $u_i$ to $u_j$ denotes that $u_j$ belongs to $u_i$'s contact list, Figure 3a. The strength value $s_{ij}^c$ of the relation $r_{ij}^c$ is calculated as follows:

$$s_{ij}^c = 1/n_i^c, \quad \text{if } u_j \text{ is in the } u_i\text{'s contact list}, \quad (1)$$

where $n_i^c = card(CL_i)$ is the number of all $u_i$'s relations derived from $u_i$'s contact list, i.e. the length of $u_i$'s contact list $CL_i$.

The relation $r_{ij}^{rc}$ from user $u_i$ to $u_j$ denotes that $u_i$ belongs to $u_j$'s contact list and is called reversed-contact relation, Figure 3b. The strength value $s_{ij}^{rc}$ of the relation $r_{ij}^{rc}$ is calculated as follows:

$$s_{ij}^{rc} = 1/n_j^c, \quad \text{if } u_i \text{ is in the } u_j\text{'s contact list}, \quad (2)$$

where $n_j^c = card(CL_j)$ is the number of all $u_j$'s relations derived from $u_j$'s contact list.

The relation $r_{ij}^{coc}$ from user $u_i$ to $u_j$ denotes that there exists another user $u_k$ that belongs to $u_i$'s contact list and $u_j$ is on the contact list of $u_k$, Figure 3c. Therefore, it represents 'contact of contact' relation, which in the literature, often refers to 'friend of the friend' type of relations. The strength value $s_{ij}^{coc}$ of the relation $r_{ij}^{coc}$ is calculated as follows:

$$s_{ij}^{coc} = n_i^{coc} / n_i^c, \quad (3)$$

where $n_i^{coc}$ is the number of different users $u_k$ on $u_i$'s contact list who simultaneously have user $u_j$ on their contact lists.

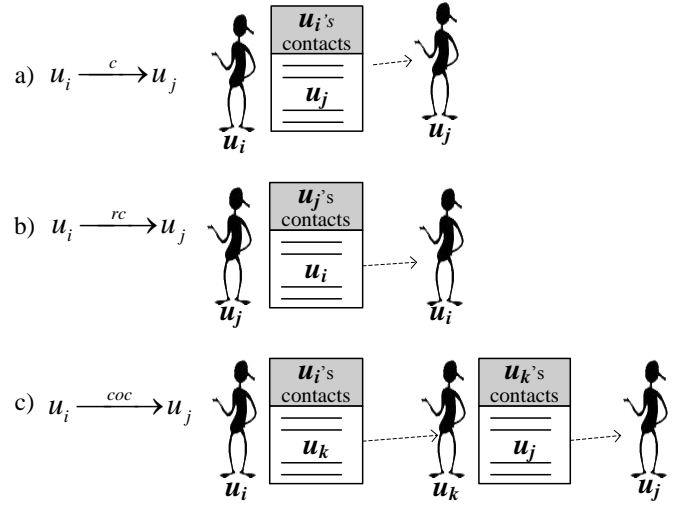

Fig. 3. Relation layers extracted from contact lists: $R^c$ (a), $R^{rc}$ (b), and $R^{coc}$ (c)

### C. Relations Based on Tags

The tag-based relation $r_{ij}^t$ between user $u_i$ and $u_j$ can be derived from information about tags they share. The general idea of extraction of tag-based relations from raw data is depicted in Figure 4.

All tags that have already been used by at least two users form the set $T$ of shared tags. The relation $r_{ij}^t$ between two users $u_i$ and $u_j$ exists if both of them have used at least one common tag to describe their photos. The strength value $s_{ij}^t$ of such relation is expressed in the following way:

$$s_{ij}^t = n_{ij}^t / n_i^t, \quad (4)$$

where $n_{ij}^t$ – the number of tags common for users $u_i$ and $u_j$; $n_i^t$ – the number of shared tags used by $u_i$.

Note that it is not important how many times tag $t_k$ was used by two users (to how many photos) but crucial is the fact that $t_k$ was shared at least once.

Tag-based relation is an object-based relation with equal roles since all users have the same role towards the picture they tag.

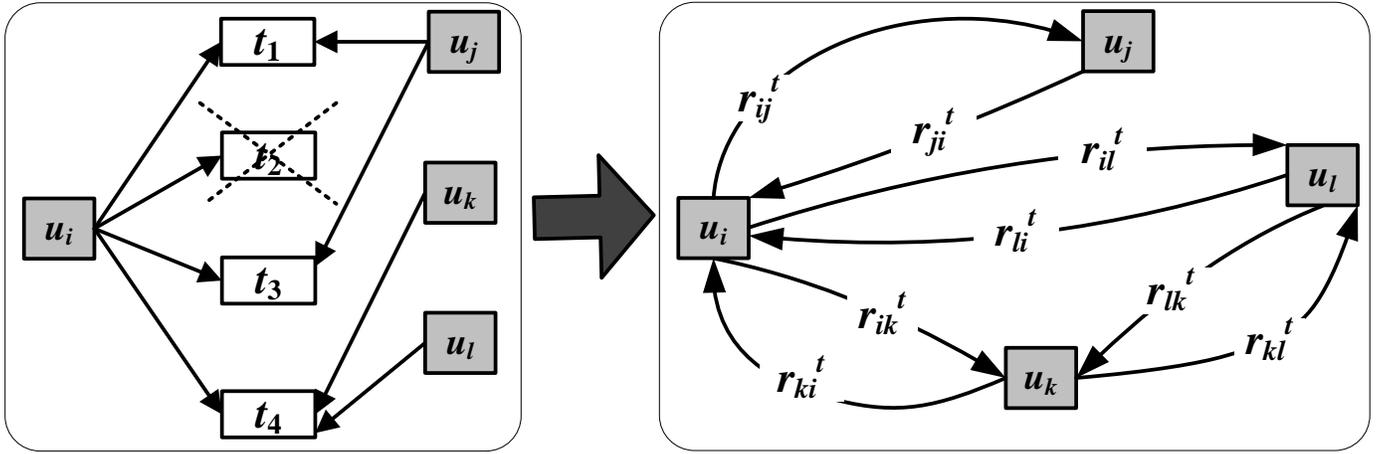

Fig. 4. Extraction of tag-based relations

### D. Relations Based on Groups

The data about groups to which users belong enable to create the relations based on groups. A group contains MOs published by a set of authors and these authors form a social group. Let $G$ be the set of all groups that consist of more than one member. The group-based relation $r_{ij}^{g}$ from user $u_i$ to $u_j$ means that users $u_i$ and $u_j$ belong to at least one common group $g_k \in G$ or to be precise there are some groups that contain photos authored by $u_i$ and simultaneously some photos published by $u_j$. The strength value $s_{ij}^{g}$ of $r_{ij}^{g}$ is:

$$s_{ij}^{g} = n_{ij}^{g} / n_{i}^{g}, \qquad (5)$$

where $n_{ij}^{g}$ – the number of groups to which belong MOs published by both users $u_i$ and $u_j$; $n_{i}^{g}$ – the number of groups containing user $u_i$'s photos.

### E. Relations Based on List of Favourites

The next three types of relations are obtained from the data about photos that have been added by some users to their favourites (Figure 5). The relation favourite-favourite $r_{ij}^{ff}$ from user $u_i$ to $u_j$ exists if both users marked at least one common photo as their favourite. The relation author-favourite $r_{ij}^{af}$ from author $u_i$ to user $u_j$ means that user $u_j$ has marked at least one $u_i$'s photo as $u_j$'s favourite. The relation $r_{ij}^{af}$ results in another relation: favourite-author $r_{ji}^{fa}$ from user $u_i$ to author $u_j$. Similarly, $r_{ij}^{ff}$ results in $r_{ji}^{ff}$. For example, when the photo $MO_m$ authored by the new user $u_i$ was marked as favourite by the first user $u_j$, then this fact creates two new relations $r_{ij}^{af}$ and $r_{ji}^{fa}$. When another user $u_k$ marks the same photo $MO_m$ then four new relations of three types are generated: $r_{ik}^{af}$, $r_{ki}^{fa}$, $r_{jk}^{ff}$, and $r_{kj}^{ff}$.

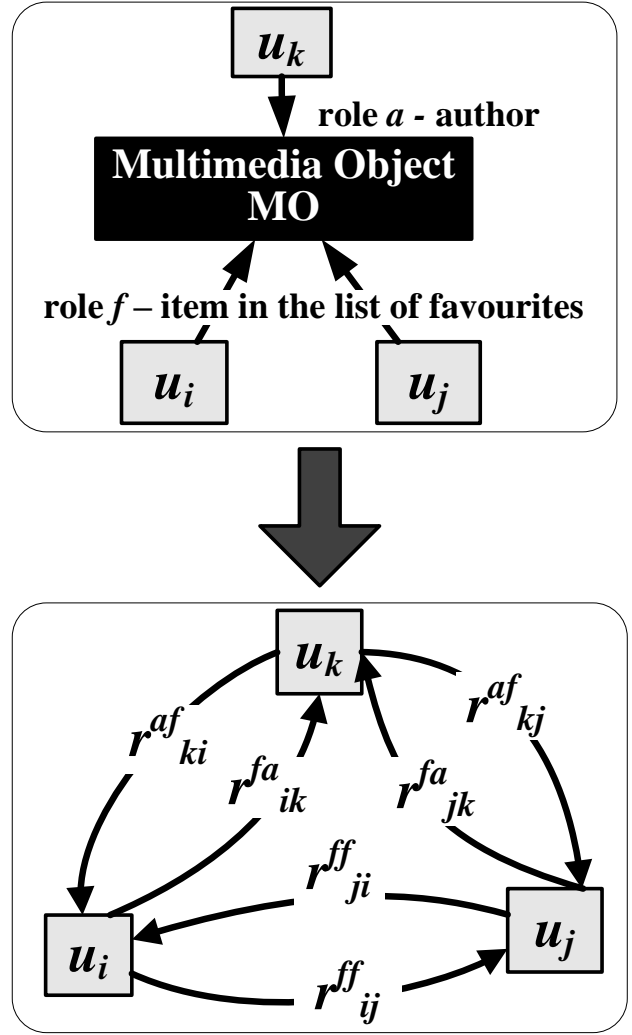

Fig. 5. Extraction of relations based on favourites

The strength value $s_{ij}^{ff}$ of relation $r_{ij}^{ff}$ is calculated as follows:

$$s_{ij}^{ff} = n_{ij}^{ff} / n_i^{f}, \qquad (6)$$

where $n_{ij}^{ff}$, $n_i^{f}$ – the number of photos marked as favourite simultaneously by user $u_i$ and $u_j$ or only by user $u_i$, respectively.

To evaluate strength value $s_{ij}^{af}$ of relation $r_{ij}^{af}$ the following formula is used:

$$s_{ij}^{af} = n_{ji}^{fa} / n_i^{a}, \qquad (7)$$

where $n_{ji}^{fa}$ – the number of photos marked as favourite by user $u_j$ and authored by $u_i$; $n_i^{a}$ – the number of all photos added by $u_i$ and marked by others as their favourite.

Finally, the formula for strength $s_{ij}^{fa}$ of relation $r_{ij}^{fa}$ is:

$$s_{ij}^{fa} = n_{ij}^{fa} / n_i^{f}, \qquad (8)$$

where $n_{ij}^{fa}$ – the number of photos marked as favourite by user $u_i$ and authored by $u_j$; $n_i^{f}$ – the total number of photos marked as favourite by user $u_i$.

Relations based on favourites are kind of object-based relation with either equal ($R^{ff}$) or different roles ($R^{af}$, $R^{fa}$), Figure 1.

### F. Relations Based on Opinions

The last three types of relations can be extracted from information about commented pictures. The relation opinion-opinion $r_{ij}^{oo}$ from user $u_i$ to $u_j$ exists if both users commented at least one common photo. The relation author-opinion $r_{ij}^{ao}$ from author $u_i$ to commentator $u_j$ exists if user $u_j$ commented at least one $u_i$'s photo. The relation opinion-author $r_{ij}^{oa}$ from commentator $u_i$ to author $u_j$ exists if user $u_i$ created opinions to at least one $u_j$'s photo.

Note that the existence of relation $r_{ij}^{ao}$ results in the reverse relation $r_{ji}^{oa}$. Favourite-based relations alike, see Sec. III.E, a single new opinion provided to the given $MO_m$ may create as many new relations as many distinct users commented this $MO_m$.

The strength values of opinion-based relations are evaluated as follows:

$$s_{ij}^{oo} = n_{ij}^{oo} / n_i^{o}, \qquad (9)$$

$$s_{ij}^{ao} = n_{ji}^{oa} / n_i^{a}, \qquad (10)$$

$$s_{ij}^{oa} = n_{ij}^{oa} / n_i^{o}, \qquad (11)$$

where $n_{ij}^{oo}$ – the number of photos commented simultaneously by user $u_i$ and $u_j$; $n_i^{o}$ – the total number of photos commented by $u_i$; $n_{ji}^{oa}$, $n_{ij}^{oa}$ – the number of photos commented by user $u_j$ and authored by $u_i$ and vice versa commented by $u_i$ and authored by $u_j$, respectively; $n_i^{a}$ – the total number of pictures authored by $u_i$ and commented by others.

Similarly to favourites, relations based on opinions are object-based relation with either equal ($R^{oo}$) or different roles ($R^{ao}$, $R^{oa}$), Figure 1.

### G. Aggregation of Layers

According to Definition 1 multidimensional social network $MSN=(U,L)$ contains the set $L$ of ties derived from data about direct intentional links between users or their shared activities. Ties (linkages) can be created as aggregation of all previously discovered relation layers. As a result, we obtain combined multidimensional social network (Figure 2). Thus, a tie $l_{ij}$ from user $u_i$ to user $u_j$ exists in the multidimensional social network, if there exists at least one relation from $u_i$ to $u_j$ of any kind. As a result, set $L$ is the sum of all relation sets identified within the system:

$$L = R^c \cup R^{rc} \cup R^{coc} \cup R^t \cup R^g \cup R^{ff} \cup R^{fa} \cup R^{af} \cup R^{oo} \cup R^{ao} \cup R^{oa}, \quad (12)$$

Tie $l_{ij}=(u_i,u_j) \in L$ reflects only the fact of connection from $u_i$ to $u_j$. Similarly to relations, we can assign real values, called strength of linkage $s_{ij}^{l}$, to each existing tie $l_{ij} \in L$ based on strengths of all component relations:

$$s_{ij}^{l} = \frac{\sum_k \alpha_k * s_{ij}^{k}}{\sum_k \alpha_k}, \qquad (13)$$

where $k$ is the index of relation layer (Figure 2); for the *Flickr* system, we have $k=1$ for $R^c$, $2 - R^{rc}$, $3 - R^{coc}$, $4 - R^t$, $5 - R^g$, $6 - R^{ff}$, $7 - R^{fa}$, $8 - R^{af}$, $9 - R^{oo}$, $10 - R^{ao}$, $11 - R^{oa}$; $\alpha_k$ – static coefficient of the $k$th layer importance; $s_{ij}^{k}$ – strength of the $k$th relation from $u_i$ to $u_j$, e.g. $s_{ij}^{1} = s_{ij}^{c}$, $s_{ij}^{2} = s_{ij}^{rc}$, ..., $s_{ij}^{11} = r_{ij}^{oa}$.

Strength of linkage aggregates all strengths from all relation levels discovered in the system. Note that values of all strengths both for relations and for ties are from the range [0;1].

Note also that one can use many different formulas for the relation strengths (Eq. 1 to 11). For example, we could incorporate the time factor into simple quantities of individual activities. In this case, each historical activity would not be counted as 1 but as $\frac{1}{\lambda^{tp}}$, where $\lambda$ is the constant and $tp$ denotes the number of fixed periods, which have passed since the time of the activity [26].

## IV. COMPARISON OF THE LAYERS IN MULTIDIMENSIONAL SOCIAL NETWORK

### A. Model for Layer Comparison

One of the aims of the paper is to compare different relation layers in the multidimensional social network based on the data gathered from *Flickr*. For that purpose nine layers $R^c$, $R^t$, $R^g$, $R^{oo}$, $R^{oa}$, $R^{ao}$, $R^{ff}$, $R^{fa}$, $R^{af}$ were extracted and analyzed.

There exist several measures to estimate the similarity between two layers. For valued relations, we can use Pearson correlation coefficient, Euclidean, Manhattan, or squared

TABLE I
STATISTICAL DATA FOR RELATION LAYERS IN *FLICKR*

| | Year | $R^c$ | $R^t$ | $R^g$ | $R^{oo}$ | $R^{oa}$ | $R^{ao}$ | $R^{ff}$ | $R^{fa}$ | $R^{af}$ | $L$ (ties) |
|---|---|---|---|---|---|---|---|---|---|---|---|
| No. of relations (% contribution in ties *L*) | 2007 | 263 (0.16%) | 3,194 (1.94%) | 163,446 (99.52%) | 288 (0.18%) | 940 (0.57%) | 461 (0.28%) | 32 (0.02%) | 156 (0.09%) | 18 (0.01%) | 164,233 (100%) |
| | 2008 | 1,464 (0.23%) | 632,330 (98.95%) | 192,396 (30.11%) | 1,278 (0.28%) | 1,278 (0.20%) | 1,257 (0.20%) | 0 (0%) | 318 (0.05%) | 318 (0.05%) | 639,033 (100%) |
| No. of non-isolated users (% of *U*) | 2007 | 191 (26%) | 361 (48%) | 679 (91%) | 106 (14%) | 264 (45%) | 135 (18%) | 31 (4%) | 143 (19%) | 16 (2%) | 745 (*U*) (100%) |
| | 2008 | 408 (43%) | 916 (97%) | 735 (78%) | 319 (34%) | 397 (42%) | 397 (42%) | 0 (0%) | 242 (26%) | 242 (26%) | 945 (*U*) (100%) |
| Average strength | 2007 | 0.73 | 0.07 | 0.07 | 0.10 | 0.28 | 0.36 | 0.97 | 0.92 | 1 | 0.008 |
| | 2008 | 0.25 | 0.08 | 0.06 | 0.04 | 0.05 | 0.05 | 0 | 0.43 | 0.58 | 0.08 |
| Strength std. deviation | 2007 | 0.30 | 0.14 | 0.11 | 0.17 | 0.32 | 0.35 | 0.12 | 0.19 | 0 | 0.016 |
| | 2008 | 0.26 | 0.11 | 0.09 | 0.098 | 0.16 | 0.12 | 0 | 0.36 | 0.35 | 0.11 |
| Avg. number of relations per user | 2007 | 1.4 | 8.8 | 240.7 | 2.7 | 3.6 | 3.4 | 1.03 | 1.1 | 1.1 | 220.5 |
| | 2008 | 3.6 | 690.3 | 261.8 | 4.0 | 3.2 | 3.2 | 0.0 | 1.3 | 1.3 | 676.2 |
| Meeting object | | N/A | Tag | Group | Commented MO | | | Favourite MO | | | Photos |
| Number of objects | 2007 | N/A | 1,718 | 13,057 | 81 | 3,112 | 1,613 | 32 | 140 | 18 | 17,905 |
| | 2008 | N/A | 481,931 | 35,826 | 2,855 | 4,787 | 4,787 | 0 | 810 | 810 | 427,914 |
| Relations per object | 2007 | N/A | 1.86 | 12.52 | 3.56 | 0.30 | 0.29 | 1 | 1.11 | 1 | 9.17 |
| | 2008 | N/A | 1.31 | 45.66 | 0.63 | 0.27 | 0.27 | 0 | 0.4 | 0.4 | 1.5 |
| Graph density | 2007 | 0.05% | 0.58% | 29.49% | 0.05% | 0.17% | 0.08% | 0.01% | 0.03% | 0.03% | 3.29% |
| | 2008 | 0.16% | 70.88% | 21.57% | 0.20% | 0.14% | 0.14% | 0% | 0.04% | 0.04% | 7.96% |
| Strength density | 2007 | 0.03% | 0.04% | 1.99% | 0.01% | 0.05% | 0.03% | 0.01% | 0.03% | 0.00% | 0.24% |
| | 2008 | 0.04% | 5.90% | 1.29% | 0.01% | 0.01% | 0.01% | 0.00% | 0.02% | 0.02% | 0.62% |

distance, whereas for binary relations, the Jaccard or Hamming measures can be utilized [17].

In further comparisons, the first enumerated measures – Pearson coefficient $p(R_1,R_2)$ is applied to calculate the similarity between two layers of relations $R_1$, $R_2$. Symbols $R_1$ and $R_2$ correspond to any two relations existing in *Flickr*. This coefficient is particularly useful when the relation between two users is directed and real valued. The range of the Pearson coefficient is [–1,1]. Value $p(R_1,R_2)=-1$ means that the corresponding relations within two layers are different while $p(R_1,R_2)= 1$ means that two layers have exactly the same links and their strengths are fully correlated.

Moreover, the layers can also be compared based on some binary measures like graph density, i.e. normalized union –
$M1 = \frac{card(R_1 \cup R_2)}{card(U)*(card(U)-1)}$, binary cosine similarity –
$M2 = \frac{card(R_1 \cap R_2)}{card(R_1)*card(R_2)}$, binary Jaccard coefficient –
$M3 = \frac{card(R_1 \cap R_2)}{card(R_1 \cup R_2)}$, or others. The values of all these binary measures from M1 to M3 belong to the range [0,1]. Note that measures M2 and M3 operate on the intersection of two relation sets and for that reason they are good indictors for the overlapping of both sets. In case of, in a sense, complementary relations like $R^{oa}$ and $R^{ao}$ or $R^{fa}$ and $R^{af}$, the measures M2 and M3 can be the sign of the common social background of both relations. It would mean that people reciprocate the interest of others, due to personal conduct rather than the semantic correlation between objects they published.

### B. Experimental Comparison of Layers
#### 1) Data Preparation

The experiment that examines the nine abovementioned relation layers over generic aggregated ties, was carried out on two *Flickr* datasets. In January 2007, the data about almost 2 million users was gathered from the *Flickr* web portal. Next, due to limited resources only top 1,000 users, who most extensively used tags, were selected together with all their associated data like contacts, groups, authored pictures, tags, comments, favourite photos. The process was repeated after a year. Therefore two datasets presents state of activity of the same *Flickr* users in January 2007 and February 2008, respectively. The reason for selecting users who utilize the greatest number of tags is that the tags are the characteristic element of all types of sharing systems and they denote how active the users are.

*2) Research Design*

Based on this data nine relation layers were extracted: $c$, $t$, $g$, $oo$, $oa$, $ao$, $ff$, $fa$, $af$ (Figure 2). Relations $R^{coc}$ and $R^{rc}$ were excluded from comparison due to technical difficulties with data gathering – it would require downloading much larger sample of the entire *Flickr* system. Users, who did not maintain any relation in any of the layers, were excluded from processing. Finally, the cardinality of the user's set ($U$) equalled 745 in 2007 and 945 in 2008. Then, using appropriate formulas Eq. 1 to 11, the strength of each relation in each layer was separately evaluated for both datasets.

Some statistics related to the proceeded data are presented in Table 1. The graph density for the $k$th layer was calculated using $\frac{card(R^k)}{card(U)*(card(U)-1)}$, whereas strength density: $\frac{\sum_{(u_i,u_j)\in L}(s_{ij}^k)}{card(U)*(card(U)-1)}$. To evaluate strength of linkage (Eq. 13), $\alpha_k=1$ was assumed for each considered layer $k$.

*3) Results and Discussion*

In 2007, $R^g$ was the largest layer. Majority of users (91%) belonged to at least one group and the number of relations in layer $R^g$ constituted 99.5% of all relations (ties) that existed within the entire multidimensional social network (Table 1). The average number of members in the group equalled 5.6 and there were 11 groups with more than 100 users. Such a big number of relations within layer $R^g$ resulted from the multiple profile of this layer. In other words, when a new user appears in the group of $N$ users, it may establish up to 2*$N$ new relations. In consequence, the average number of relations that one person maintained within $R^g$ in 2007 was over 240 and the graph density was almost 30%. However, this was not valid to the same extent for other object-based relations with equal roles like tag-based ($R^t$), favourite-favourite ($R^{ff}$), and opinion-opinion ($R^{oo}$). The relatively high number of relations per object: 12.5 for $R^g$, 3.56 for $R^{oo}$, and 1.86 for $R^t$ resulted in small values of strength, in average below 0.1.

In 2008, $R^g$ was not the biggest layer any more. Admittedly, still majority of users (77%) belonged to at least one group but $R^g$ was the component of only 30% of all ties. The average number of members in the group increased 16 times up to 93.2 and there were 1,811 groups with more than 100 users in 2008. Moreover, there existed 47 groups with more than 500 users.

In 2008, the average number of relations per user in $R^g$ was over 262 (9% growth) and graph density was 21.5% (36% decrease).

The snapshot of the described social network from 2008 exposed that $R^t$ had become the most dense and strongest layer within the entire multidimensional social network – it was included in 99% of all ties extracted. As compared to 2007, it can be observed nearly 200 times higher number of relations in $R^t$ layer in 2008 as well as the density on the level of almost 71% revealed significant growth.

The experiments revealed that "folksonomy" concept (tagging of photos) has been accepted by most of users and the acceptance rate has risen year by year. Hence, tags have become the most significant meeting object between users – growth from 48% to 97% of users participating in $R^t$. The number of used tags increased 280 times up to over 480 thousand!

The interesting fact is that users are likely to maintain only few contacts and therefore their relations in layer $R^c$ are relatively strong: in 2007 only 1.4 relations per user with average strength – 0.73 and in 2008 – 3.6 relations per user with average strength – 0.25.

In case of layers $R^c$, $R^g$, $R^{oo}$, $R^{fa}$, $R^{af}$ the change of average relations per user between 2007 and 2008 is inversely proportional to the change of average strength, obviously with varies proportion factors. Contrary situation can be observed in $R^t$ layer, where with growth of average relations per user rises the average strength. For the rest of layers ($R^{oa}$, $R^{ao}$), decrease of average relations per user is accompanied by decrease of average strength.

The average strength of relations for $R^{fa}$ and $R^{af}$ is very high (over 0.9 in 2007 and still very high – over 0.4 in 2008), which can be interpreted as a single user tendency to mark as favourite MOs of only very few other users they feel to be close to. This was also valid for $R^{oa}$ and $R^{ao}$ in 2007, nevertheless their strengths were not so big. In 2008, $R^{oa}$ and $R^{ao}$ were significantly weaker. Overall, it probably means that people, who add to their favourites or comment MOs of another users, utilize for this purpose the acquaintance with that user rather than semantic relationships between MOs. Thus, the basis of $R^{fa}$ and $R^{af}$ as well as $R^{oa}$ and $R^{ao}$ is more social than semantic. This also effects direct intentional relations like contact-based $R^c$ – the average strength was 0.73 in 2007 and 0.25 in 2008. The usage of more semantic approach in user activities would cause dispersal and downgrading of relation strengths. This can be observed for tag-based and all opinion-opinion based relations – the average strength is below 0.1. Hence, $R^t$ and $R^{oo}$ are more semantic based in opposite to others.

The layers have been also compared by means of following measures: binary cosine similarity (M1), binary Jaccard coefficient (M2), and Pearson coefficient (M3). M1 and M2 are binary measures, whereas Pearson coefficient respects real values of strength, Figure 6.

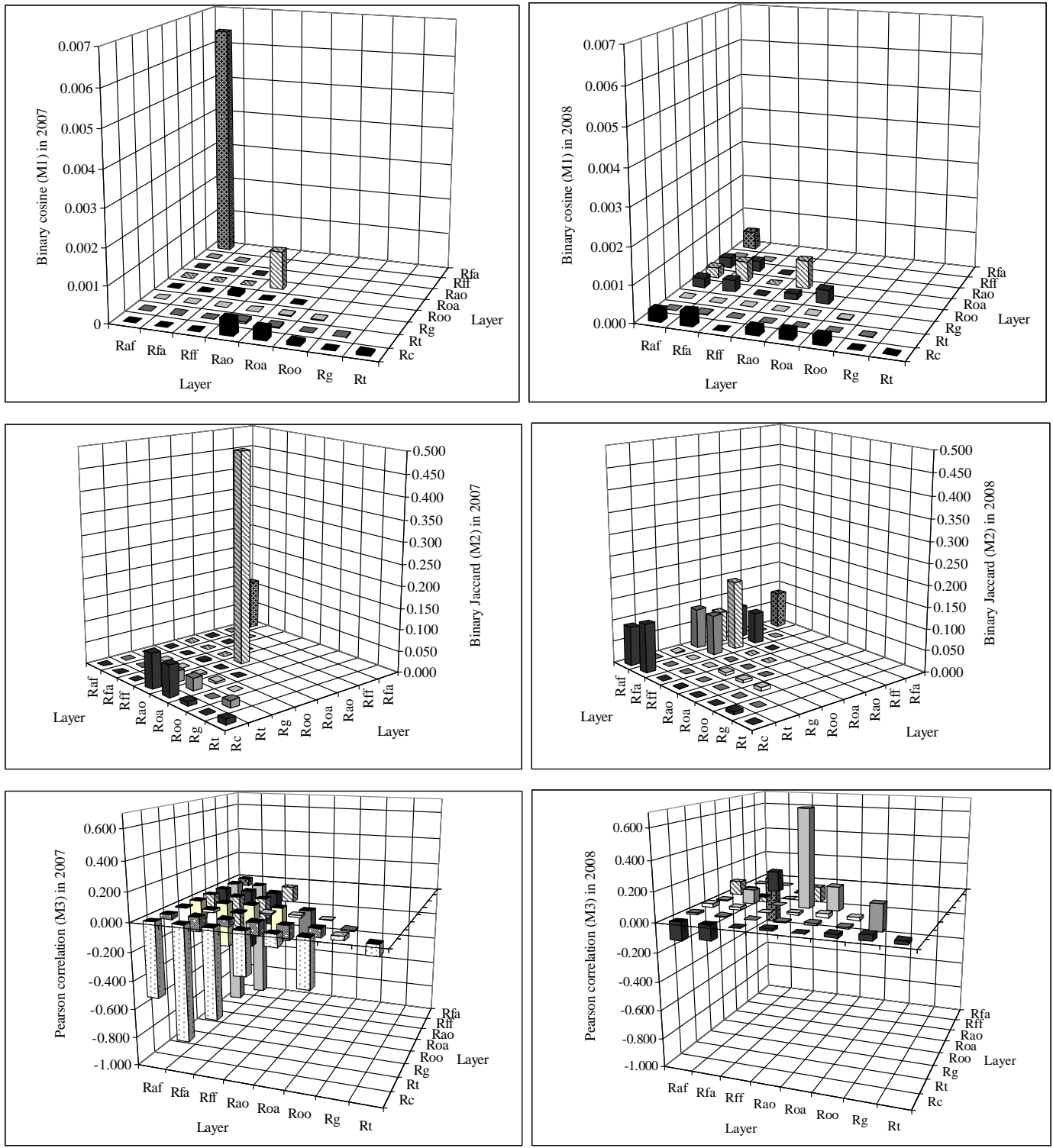

Fig. 6. Similarity between relation layers: binary cosine M1, binary Jaccard M2, Pearson correlation M3, separately as of 2007 and 2008

In 2007, the highest Pearson correlation M3 was between $R^{oa}$ and $R^{ao}$ (0.1) while in 2008, between $R^{ao}$ and $R^{oo}$ layers (0.682), Figure 6.

On the other hand, layers $R^{ff}$, $R^{af}$, $R^{fa}$ were strongly divergent according to Pearson coefficient. Generally, two users who created relation $r^{ff}$ by adding the same MOs to their favourite, established neither $r^{af}$ nor $r^{fa}$ relation between them. In 2008, there were no $r^{ff}$ relations at all. It can stand for basically social profile of relations $r^{af}$ and $r^{fa}$.

The general conclusion is that pictures are added to favourites because of their authors rather than their subject. This rather social inspiration of activities based on favourites are additionally confirmed by the mutuality of relations $R^{fa}$ and $R^{af}$ – the highest value of binary cosine measure

M1($R^{fa}$,$R^{af}$)=0.0064, the second highest M2($R^{fa}$,$R^{af}$)=0.12 in 2007 as well as the third highest M1($R^{fa}$,$R^{af}$)=0.00049 in 2008. Values of M1 are the highest for the layers $R^{fa}$ and $R^{af}$ in both years. Even greater social involvement could be observed between $R^{ao}$ and $R^{oa}$ in 2007: the undisputed highest value of M2($R^{ao}$,$R^{oa}$)=0.49, the highest Pearson correlation 0.1, and the second highest M1($R^{ao}$,$R^{oa}$)=0.0011. Note that all others values of M1 in 2007 were below 0.0005 and for M2 below 0.085. Charts of M1 and M2 as well as to a large extent of M3 show that $R^{fa}$ and $R^{af}$ are correlated with the other layers except $R^c$ and $R^t$ in 2008 (Figure 6).

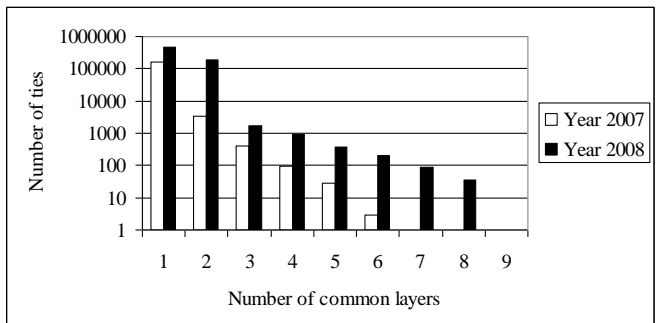

Fig. 7. Number of ties in relation to the number of common layers

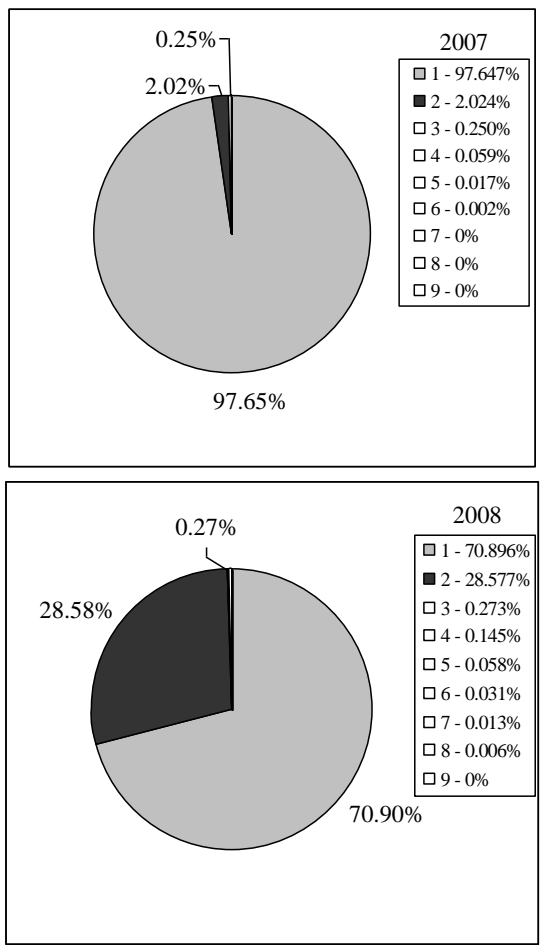

Fig. 8. Distribution of number of common layers in entire multidimensional social network

It generally means that if user $u_i$ adds to favourites or comments pictures of user $u_j$ then in many cases user $u_j$ reciprocates by adding $u_i$'s photos to their favourites or comments them, respectively. Similarly, if user $u_i$ and $u_j$ "meet each other" while commenting somebody's else photo, they are also likely to comment their photos each other.

Overall, tag-based layer ($R^t$) reflect semantic relationships between users whereas the other layers, especially favourite-based and opinion-based ($R^{fa}$, $R^{af}$, $R^{oa}$, $R^{ao}$ and $R^{oo}$) have more social profile.

The relations in separate layers complemented one another in 2007 – the number of relations common for two or more layers is relatively small – 4,026 relations, less than 2.4% of total (Figure 7 and 8), see also intersection-based measures M1 and M2 in Figure 6. This trend was reversed in 2008 on the grounds of enormous increase of $R^t$ that met with $R^c$ while the other layers remained independent.

## V. Social Recommendations within Multidimensional Social Network

The main idea of recommendations in the multimedia sharing system is to make use of relations from the multidimensional social network to suggest to active given user some others potentially interesting ones. The entire recommendation process is presented in Figure 9.

In the first step, multidimensional social network MSN is created and continuously updated based on the data about user activities available in MSS, including comments, favourites, contact lists or groups, etc., see Sec. III. The values of relationship strengths $s^k_{ij}$ are calculated for each connection from user $u_i$ to $u_j$ existing in $k$th layer, according to formulas Eq. 1 to 11.

Additionally, two kinds of weights are maintained by the system: system weights and personal weights. Both are separately evaluated for each layer $k$ in MSN. They reflect seither general or individual importance of a given MSN layer in the recommendation process.

The system weight $w_k^{sys}$ for layer $k$ is the aggregation of all personal weights for layer $k$: the sum of all personal weights divided by the number of users in a given layer. Usually, system weights need to be periodically updated, e.g. once a day.

Personal weight $w_{ki}^{usr}$ of layer $k$ reflects the current usefulness of layer $k$ for user $u_i$. Both system and personal weights are in the range [0;1] but the sum of all $w_{ki}^{usr}$ for user $u_i$ equals 1. For a new user $u_i$, at the beginning, the formula $w_{ki}^{usr}=w_k^{sys}$ is applied. All personal weights for user $u_i$ are updated according to $u_i$'s activities related to recommended people like browsing profiles of others, adding others to $u_i$'s contact list, comments to MOs published by recommended users, etc. In this way, the system monitors usability of suggestions provided. In the experimental environment, users were requested to rate the presented recommendations and these rates were used as the feedback from user's activities, see Sec. VI.

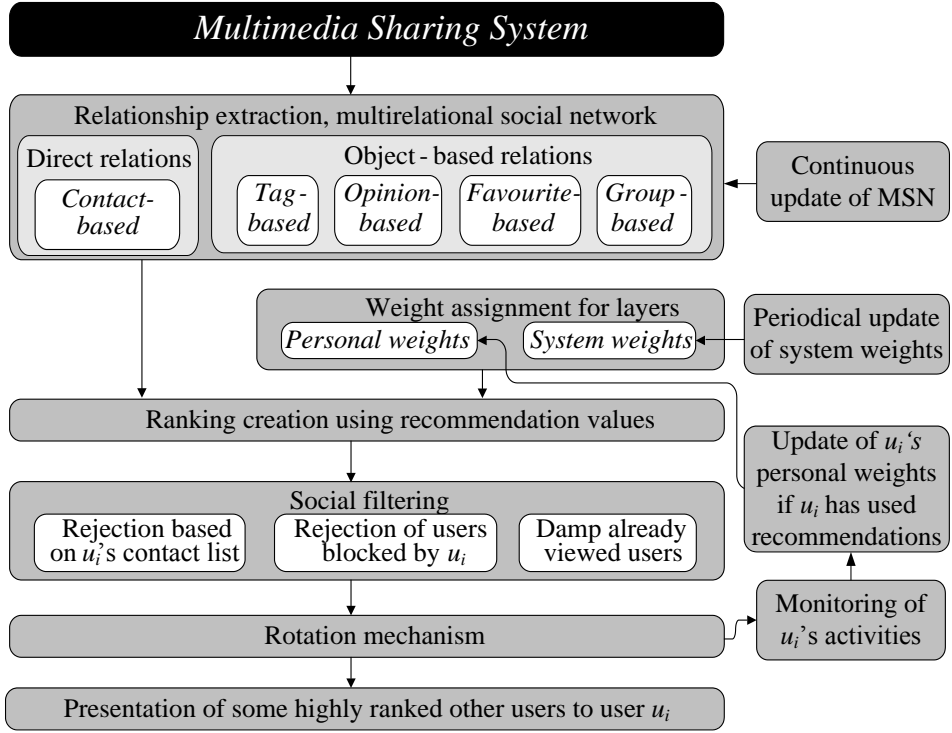

Fig. 9. Recommendation of humans in MSS for user *k* using multidimensional social network

Relation strengths $s^k_{ij}$ can be seen as the degree to which two users are similar to each other. They are, together with system and personal weights, used to calculate the recommendation values $v_{ij}$ for the current user $u_i$ related to the other users $u_j$, as the aggregations of similarities from all *l* layers, as follows:

$$v_{ij} = \sum_{k=1}^{l} \frac{(w_k^{sys} + w_{ki}^{usr}) \cdot s_{ij}^k}{\max_k \left(s_{ij}^k\right)}, \quad (14)$$

Afterwards, the recommendation values $v_{ij}$ are utilized to create the ranking list for user $u_i$. Such list contains *N* top users $u_j$ with the greatest value of $v_{ij}$. However, some users $u_j$ are removed from the list during the social filtering process. Its goal is to prevent from recommendation of users that have already been in the $u_i$'s contact list or have been blocked by $u_i$. Besides, the recommendation values of users who have already been viewed by $u_i$ are reduced. A rotation mechanism is applied to the remaining list so that the recommendation list changes with every user request to the system, see [21] for more details. Finally, some top ranked users from the processed list are presented to the current user $u_i$. Next user $u_i$ can watch profile of the suggested persons and perform some other actions related to them.

After presentation, the system strictly monitors consecutive activities of user $u_i$ related to the recommended users $u_j$. It includes especially viewing the $u_j$'s profile, commenting $u_j$'s photos, adding them to $u_i$'s favourite or even putting $u_j$ to $u_i$'s contact list. These interactions are the basis to establish a new relation $u_i \rightarrow u_j$ in one or more MSN layers. However, the level of interest of user $u_i$ directed to $u_j$ reflected by $u_i$'s activities can be lower (just viewing $u_j$'s profile) or greater (directly appending to $u_i$'s contact list). Hence, each type of activity *n* possesses its own importance $a_n \in [0;1]$. For example, viewing profile of the recommended person (activity *n*=1) can possess the value $a_1$=0.1, whereas extension of the contact list with the user suggested by the system (action *n*=5) may be much more meaningful $a_5$=1.

Next, based on this feedback, $u_i$'s personal weights $w_{ki}^{usr}$ are adapted after each $u_i$'s activity *n* relevant to user $u_j$, separately for each layer *k*, as follows:

$$w_{ki}^{usr(new)} \overset{u_i\text{'s action } n \text{ towards } u_j}{=} \frac{w_{ki}^{usr(old)} \cdot (1+\varepsilon) + c_{ij}^k \cdot \left(a_n - w_{ki}^{usr(old)}\right)}{\sum_{m=1}^{l} \left(w_{mi}^{usr(old)} + c_{ij}^m \cdot \left(a_n - w_{mi}^{usr(old)}\right)\right)}, \quad (15)$$

where $\varepsilon$ is a very small constant; $c_{ij}^k \in [0;1]$ is the normalized contribution of the *k*th layer (among all layers) for the recommendation of user $u_j$ to user $u_i$.

The value of $c_{ij}^k$ is calculated in the following way:

$$c_{ij}^k = \frac{s_{ij}^k}{\sum_{m=1}^{l} \left(s_{ij}^m\right)}, \quad (16)$$

Eq. 15 ensures that the value of $w_{ki}^{usr(new)}$ is from the range [0;1]. It takes into account the global importance of particular kinds of relations in the entire MSN. The more importance gains layer *k* for user $u_i$, the stronger relationship between $u_i$ and $u_j$ in this layer *k* (greater value of $s_{ij}^k$ in Eq. 16). It means that user $u_i$ was successfully attracted with user $u_j$ based on stronger relation in the certain layer *k* in MSN, so layer *k* should benefit.

Due to efficiency issues, both weight updates (Eq. 15 and 16) as well as revision of ranking lists (Eq. 14) should not be performed after each user activity but lunched offline and periodically repeated, e.g. once a day.

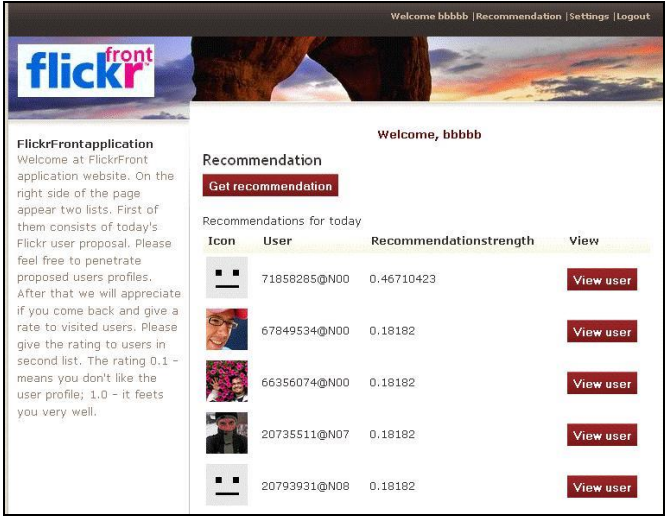

Fig. 10. *FlickrFront* system for recommendation of humans in MSS

## VI. EXPERIMENTS ON SOCIAL RECOMMENDER SYSTEM

### A. Data Preparation

The experiments have been carried out based on the online *FlickrFront* framework, which has been specially implemented for this purpose, Figure 10. During the experiments, 21,640 user profiles were downloaded from *Flickr* to prepare recommendations for eight volunteers by means of all eleven layers from the multidimensional social network, Figure 10.

### B. Research Design

The experiment consisted of two main stages with two separate recommendation lists. The first one contained initial suggestions computed using equal values of personal weight for layers, i.e. for each layer $k$ and each user $u_i$: $w_{ki}^{usr}=1/11$. Next, users were asked to rate the provided suggestions.

In the second phase, personal weights were adjusted separately for each user using adaptation mechanism, Eq. 15 and layer contributions were applied after the first lists were rated, Eq. 16. The rates provided by users replaced the monitored user activities $a_n$ in Eq. 15. Additionally, the rates were used to assess the usefulness of recommendations. Afterwards, recommendation lists were recalculated and again presented to users, two weeks later. Obviously, persons suggested during the first stage, were excluded from the second list.

### C. Results and Discussion

Users have generally rated higher the recommendations provided in the second list (after adaptation), in average 8% better, which proves higher usability of recommendations with adaptation mechanism, Figure 11.

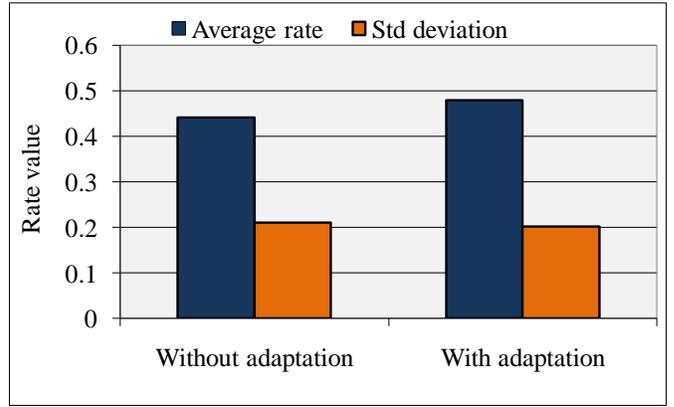

Fig. 11 Average user ratings for layers in MSN

Analysis of personal weight values after adaptation (second stage) revealed that the social layer based on indirect reciprocal contact list $R^{coc}$ and author-opinion $R^{ao}$ gained in their contribution (average $w_{ki}^{usr}$) much after adaptation, by 220% and 65% respectively, Figure 12. Moreover, it referred to all volunteers participating in the experiment in case of $R^{coc}$ and most for $R^{ao}$, Figure 13. Tag-based layer $R^t$ increased in average by 8%, whereas the other layers lost in their importance. The least significant layers were $R^{oa}$, $R^{ff}$, $R^{af}$; they droped in average by -59% to -66%, Figure 12. The distribution of changes in personal weights differed for particular individuals, Figure 13, although the general trends are rather clear, Figure 12.

These results have confirmed that the proposed method of weight adaptation (Eq. 15) increased the user satisfaction (rise in rates). This has been achieved by highlighting strong social components ($R^{cac}$) at expense of some semantic relationships like favourite-favourite ($R^{ff}$). The layer opinion-author $R^{oa}$ lost and author-opinion $R^{ao}$ gained because the former reflect relationships to authors of MOs that have been commented by a given users. These authors are not so attractive compared to those who have commended photos delivered by the current user. Hence, people tend to be interested in other people who reviewed their achievements.

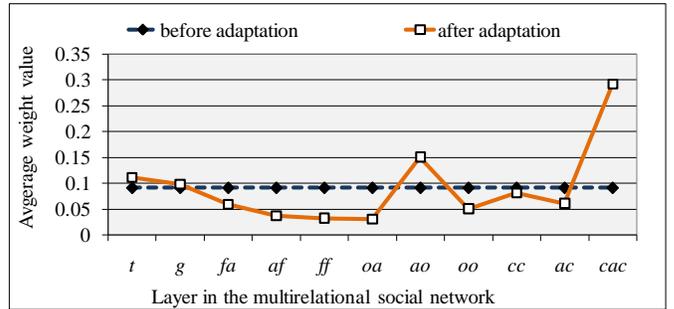

Fig. 12 Average user weights for layers in MSN

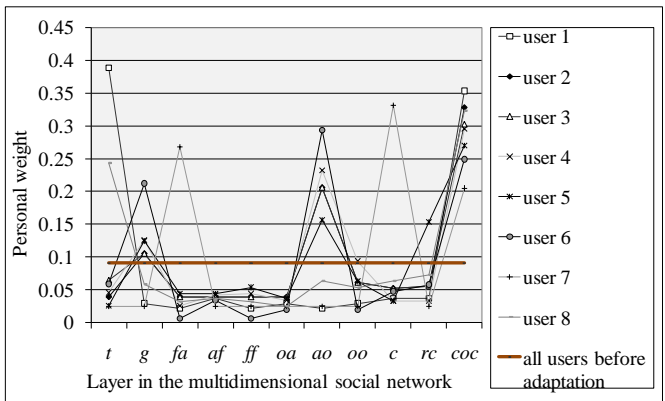

Fig. 13 User weights before and after adaptation

## VII. CONCLUSIONS

The *Flickr* users form a social network of people with common interests and activities. The members of this social network can be related either directly or indirectly through an external object like commented picture, group or tag they share. Based on these different kinds of connections, many separate relation layers can be identified in the multidimensional social network MSN. These layers usually complement one another. Moreover, more than one relation in more than one layer can be created for the single user activity, e.g. a new comment on the single picture may result in some new relations: between commentator and photo's author (opinion-author and author-opinion relations) as well as between the commentator and all other commentators of this photo (many opinion-opinion relations). The multidimensional social network that aggregates all existing layers provides a comprehensive view onto relationships between users in MSS. It merges both semantic and social backgrounds of user activities. Semantic inspiration of users refer especially tag-based relations whereas opinion- and favourite-based relations that link authors with others interested in their photos reflect more social motivations. This duality shows that social networks in complex multimedia publishing systems should be considered using many dimensions.

The spanned of over a year research revealed that tag-based relations (folksonomy) more and more dominate the multidimensional social network created within the online publishing system. Overall, the multidimensional social network becomes more affluent in its component layers. As a result, users are more and more related to others through different dimensions – the number of ties linking the same set of users increased almost four times year by year.

The social network, which can be extracted from user activities, can support other cooperative actions of users like collaborative Information Retrieval or metadata management. It also facilities trust management between its members, targeted marketing [21] and especially recommender systems.

The new recommendation method presented in the paper provides suggestions of other users in the multimedia sharing system based on the knowledge discovered in multidimensional social network. The presented framework takes into account all users' activities stored in separate layers of MSN. The system and personal weights that are assigned independently to each layer make the recommendation process personalized. Additionally, the system is adaptive due to personal weights that are adaptively recalculated when the user utilizes the recommendations.

The vast amount of calculations results in problems with efficiency as the whole process is performed online. In order to address this issue, some tasks can be performed offline and periodically repeated, e.g. the creation of the lists and storing only *n* most similar users to the given one.

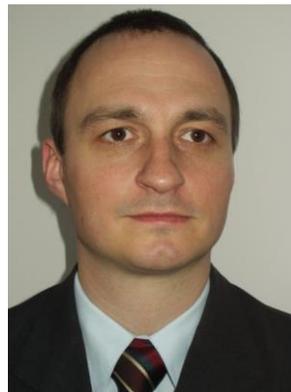

**Przemysław Kazienko**, Ph.D. was born in 1966 in Poland. He received the M.Sc. and Ph.D. degrees in computer science both from Wrocław University of Technology, Poland, in 1991 and 2000, respectively. He obtained his habilitation degree from Silesian University of Technology, Poland, in 2009.

Recently, he serves as a professor of Wroclaw University of Technology at the Institute of Informatics, Poland. He was also a research fellow at Intelligent Systems Research Centre, British Telecom, UK in 2008. For several years, he held the position of the deputy director for development at Institute of Applied Informatics.

He was a co-chair of international workshops RAAWS'05, RAAWS'06, MMAML 2010, MMAML 2011, SNAA 2011 and a Guest Editor of New Generation Computing and International Journal of Computer Science & Applications. He regularly serves as a member of international programme committee and the reviewer for scientific conferences and prestige international journals. He is a member of Editorial Board of International Journal of Knowledge Society Research and International Journal of Human Capital and Information Technology Professionals. He has authored over 110 scholarly and research articles on a variety of areas related to data mining, recommender systems, social networks, knowledge management, hybrid systems, collaborative systems, Information Retrieval, data security, and XML. He also initialized and led over 20 projects chiefly in cooperation with commercial companies, including large international corporations.

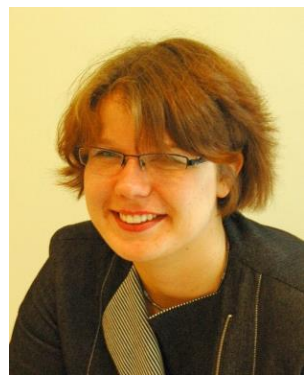

**Katarzyna Musiał** was born in 1982 in Poland. She received her M.Sc. degree in Computer Science from the Wroclaw University of Technology, Poland in 2006. In the same year she received her second MSc degree in Software Engineering from the Blekinge Institute of Technology,

Sweden. She obtained Ph.D. in 2009 from the Institute of Informatics, Wroclaw University of Technology, Poland. She is a Lecturer in Informatics at Bournemouth University, UK.

Katarzyna Musiał focused her Ph.D. thesis on the calculation of individual's social position in the virtual social network. She is interested especially in complex social networks and dynamics and evolution of complex networked systems. She is one of the founder members of the 'Social Network Group' at WUT (SNG@WUT) established in October 2006.

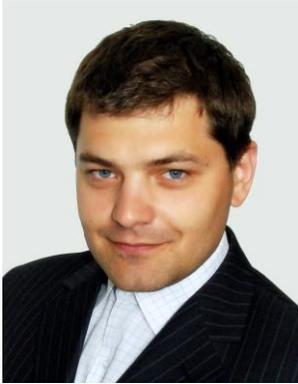

**Tomasz Kajdanowicz** was born in Poland in 1983. He received the M.Sc. degree from Wroclaw University of Technology, Poland in 2008.

He is currently a PhD student at Wroclaw University of Technology and simultaneously a consultant at Hewlett Packard, Poland.

His research areas focus on social network analysis and hybrid information systems as well as their applications, especially in the industry. While participating in multiple research and development projects, he collaborates with leading financial enterprises in Poland. He authored a dozen of scientific papers and articles.